# A Multi-Technique Study of $CO_2$ Adsorption on $Fe_3O_4$ Magnetite

Revised 12/14/2016  10:12:00

Jiri Pavelec, Jan Hulva, Daniel Halwidl, Roland Bliem, Oscar Gamba, Zdenek Jakub, Florian Brunbauer, Michael Schmid, Ulrike Diebold and Gareth S Parkinson

*Institute of Applied Physics, TU Wien, Wiedner Hauptstrasse 8-10, 1050 Wien, Austria*
*e-mail address: parkinson@ iap.tuwien.ac.at*

The adsorption of $CO_2$ on the $Fe_3O_4(001)$-$(\sqrt{2}\times\sqrt{2})R45°$ surface was studied experimentally using temperature programmed desorption (TPD), photoelectron spectroscopies (UPS and XPS), and scanning tunneling microscopy (STM). $CO_2$ binds most strongly at defects related to $Fe^{2+}$ including antiphase domain boundaries in the surface reconstruction and above incorporated Fe interstitials. At higher coverages, $CO_2$ adsorbs at fivefold-coordinated $Fe^{3+}$ sites with a binding energy of 0.4 eV. Above a coverage of 4 molecules per $(\sqrt{2}\times\sqrt{2})R45°$ unit cell, further adsorption results in a compression of the first monolayer up to a density approaching that of a $CO_2$ ice layer. Surprisingly, desorption of the second monolayer occurs at a lower temperature (≈ 84 K) than $CO_2$ multilayers (≈ 88 K), suggestive of a metastable phase or diffusion-limited island growth. The paper also discusses design considerations for a vacuum system optimized to study the surface chemistry of metal oxide single crystals, including the calibration and characterisation of a molecular beam source for quantitative TPD measurements.

## 1. INTRODUCTION

$CO_2$ is one of the most common components in the atmosphere of planets and interstellar dust, which makes understanding both the gaseous and solid phases important for astrophysical research.[1-3] On Earth, emissions of $CO_2$ into the atmosphere are rising, and there is a growing effort to develop carbon-capture and storage technologies ($CO_2$ sequestration) to mitigate global warming. As such there is much interest in the interaction of $CO_2$ with components of the environment including water and Earth abundant minerals.[4] The adsorption and activation of $CO_2$ is also important in catalysis, including reactions such as CO oxidation, water-gas shift and Fischer-Tropsch synthesis, where iron oxide materials are often used as catalysts, or as a support for metal nanoparticles.[5] Finally, $CO_2$ is often utilized as a probe of the basicity of metal oxide surfaces.[6]

Fundamental investigations of $CO_2$ adsorption on well-characterised metal-oxide surfaces are scarce. The interaction is generally stronger than on clean metal surfaces,[7] but ranges from physisorption on clean $TiO_2$,[8] $ZnO$,[9] and $MgO$,[10] to carbonate formation on $CaO$ [11] and $Cr_2O_3$ [12] surfaces. Surface defects such as oxygen vacancies tend to bind $CO_2$ more strongly than the regular surface,[13] and $CO_2$ adsorption can therefore be used as a quantitative probe of the defect concentration.[14] To date there are no investigations of $CO_2$ adsorption on well-defined iron-oxide surfaces, despite the important role of these materials in both geochemistry and catalysis. Work on polycrystalline $Fe_3O_4$,[15] $Fe_3O_4$ nanoparticles,[16] and $FeO_x$ nanoclusters on graphite[17] suggest however, that both physisorption and carbonate formation can occur, with stronger binding linked to the presence of $Fe^{2+}$ cations.[18, 19] Very recent DFT calculations based on the $Fe_3O_4(111)$ surface suggest that $CO_2$ chemisorption can occur at undercoordinated oxygen sites, and that this surface can activate $CO_2$ for hydrogenation.[20]

In this paper, we study the adsorption of $CO_2$ on the $Fe_3O_4(001)$ surface utilizing an experimental ultrahigh-vacuum (UHV) setup optimized to study the surface chemistry of single-crystal metal-oxide samples. The paper begins with a description of the new vacuum system, with a focus on how we combine an effusive molecular beam (MB) source and a special sample mount to perform quantitative temperature programmed desorption (TPD) measurements of bulk oxide single crystals. Then, we utilize TPD data along with photoelectron spectroscopies and scanning tunneling microscopy (STM) data to show that $CO_2$ adsorbs on $Fe^{2+}$-related defects on $Fe_3O_4(001)$ initially, and then forms a physisorbed monolayer with molecules adsorbed on regular, undercoordinated $Fe^{3+}$ surface sites. Additional $CO_2$ molecules initially compress the monolayer before a complete second layer is formed. Surprisingly, this second monolayer is less strongly bound than multilayer $CO_2$ ice.

## 2. DESCRIPTION OF THE UHV SYSTEM

The experiments were performed in a newly constructed UHV system that was designed to study the surface chemistry of single-crystal metal-oxide samples and oxide-supported nanoparticle systems. A schematic view of the setup is shown in Fig. 1. The UHV chamber itself is constructed from μ-metal and achieves a base pressure of $5\times10^{-11}$ mbar. The sample mount is attached to a Janis ST-400 flow cryostat on the central axis of the chamber, and can be rotated and moved between two levels using a Thermionics EMX xyz manipulator. The upper level (Fig. 1b) is primarily for spectroscopy, and utilizes a SPECS Phoibos 150 energy analyser with nine channeltrons for charged-particle detection. X-ray photoelectron spectroscopy (XPS) is performed with a SPECS FOCUS 500 monochromated x-ray source (Al Kα or Ag Lα anode), ultraviolet photoelectron spectroscopy (UPS) utilizes a SPECS UVS 10/35 source with both He I and He II discharge, and low-energy ion scattering (LEIS) is performed using a SPECS IQE 12/38 ion source. This ion source is also used to generate the $Ne^+$ ions required for sample sputtering. We use Ne rather than Ar because the latter would condense on the cryostat. Finally, a simple tube doser is installed to direct $O_2$ to the sample surface during oxidative annealing cycles, although the MB source can also be used for this purpose.

The lower level (Fig. 1c) has a low-current low-energy electron diffraction multichannel plate (MCP-LEED) optics (Omicron) and several ports for metal evaporators or Knudsen cells as well as a quartz-crystal microbalance (QCM). The primary function of the lower level, however, is to conduct TPD experiments. Included for this purpose is a quadrupole mass spectrometer (HIDEN HAL 3F PIC) and a home-built molecular beam (MB) source and beam monitor (BM). For TPD experiments, the mass spectrometer is moved on a linear motion to a position with its front end 11 mm from the sample surface; this increases the sensitivity to minor reaction products. For shorter separations than this, desorbing molecules were found to reflect back from the mass spec yielding ghost peaks in TPD data. To further reduce this effect, the mass spec shielding was removed. During TPD measurements the sample is biased to -70 V to prevent electrons from the mass spec filament reaching the surface. In what follows we explain the rationale behind the design of the sample mount, MB and beam monitor, since these components were optimized for the study of metal-oxide single crystals.

### 2.1 TPD of Metal-Oxide Single Crystals

A major design goal was to perform high-quality (quantitative) TPD studies on metal oxide single crystals. It is not straightforward to achieve reproducible thermal contact and temperature measurement on such samples, and we followed the approach described in the work of Kay et al.[21-23] and Kimmel et al.[24, 25] The metal oxide sample is mounted on a metal backplate, and the temperature is measured by K-type thermocouple spot-welded to the backplate. The temperature at the sample surface is calibrated by multilayer desorption, as described by Menzel et al.[26] In our setup, a base temperature of ≈30 K is achieved at the sample surface. The sample is heated by resistive heating of the sample plate, and good thermal contact is assured by pressing the sample to the backplate using Ta clips (see Fig. 2a). Temperatures up to 1200 K can be achieved. It is important to note, however, that the clamps and backplate introduce significant additional surface area close to the sample, from which unintended desorption could occur and complicate data analysis. Thus, it is crucial that only the sample surface is exposed to reactant gases, which is achieved using the MB.

**2.2 Molecular Beam**

For our system, we required a uniform coverage of adsorbates over a well-defined area, with minimal exposure outside the intended area. Full details of the MB can be found in the supplement and the work of Halwidl[27] but briefly, a MB of diameter 3.5 mm at the sample is produced by expansion of gas from a reservoir (pressure typically of the order mbar, and measured by a capacitance gauge) through a thin-walled orifice with an effective diameter of 37.9 ± 0.4 μm and thickness of 20 μm. The effective diameter quoted above was estimated using a calibrated SEM image (Fig. S1b). An alternative measurement using the flow through method of Yates et al.,[28] determined an effective diameter of 37 ± 6 μm. The beam passes through two stages of differential pumping over a distance of 90 mm, and entry into the chamber is controlled by an electromagnetic shutter covering the aperture between the first and second pumping stage. The thin-walled orifice, the aperture between the differential pumping stages, and the beam-defining exit aperture are mounted to a rigid body to achieve precise alignment and mechanical stability. The exit aperture (diameter 2.0 mm) is placed at a distance of 39 mm to the sample. This close proximity results in a narrow beam penumbra. With this setup, calculations of the beam shape (see supplement for details) predict that 97.7% of the molecules lie within the beam core, 2.2% inside a penumbra of width 38 μm, and a further 0.1% inside a penumbral region of width 0.75 mm. Only ≈0.01 % contribute to the background signal (see Fig. 2b). In the effusive regime, the final pressure in the core of the beam depends linearly on the reservoir pressure and is of the order $10^{-8}$ mbar. The relatively slow deposition rate ensures that submonolayer exposures can be achieved with high precision. Crucially for our purposes, the rate at which molecules are deposited to the surface can be calculated directly from the reservoir pressure and beam geometry. We demonstrate in this paper that the calculated values are in excellent agreement with STM and XPS measurements. If higher pressures are required, the core pressure can be increased up to $10^{-5}$ mbar by increasing the reservoir pressure. However, this occurs at the expense of linearity of response and induces a minor smearing of the beam profile.

**2.3 Molecular Beam Profile Measurements**

To check the profile of the MB we constructed a beam monitor (BM). Briefly, the BM is an accumulation detector, inspired by the design of Libuda et al.[29]. In our case, a 0.5 mm orifice provides entrance to a closed volume containing a Granville-Phillips micro-ion gauge. To minimise the closed volume, we placed the ion gauge within the vacuum chamber. This improves the response of the detector, but heat generated by the ion gauge necessitated cooling via a connection to the exterior that is achieved using a copper rod. A plot of the measured beam core pressure versus the reservoir pressure (Fig. 2c) exhibits the expected linear dependence over the majority of the required range. However, the precision of the ion gauge (≈15%) is insufficient for calibration of the beam flux for our TPD experiments. Measurements of the beam profile are included in Fig. S3, and agree with the calculations within the limit of the MBM orifice size. In Fig. 2a, we show an alternative visualization of the beam shape obtained by dosing water in the effusive regime onto the sample surface at 100 K until the spot became thick enough to be visible to the eye.

## 3. $CO_2$ ON $Fe_3O_4$ - EXPERIMENTAL

A natural $Fe_3O_4$(001) single crystal with miscut precision < 0.1° and dimensions 6×6×1 mm was purchased from Surface Preparation Laboratory. The sample exhibited a sharp Verwey transition at 124 K, an indicator of excellent stoichiometry and purity.[5] The crystal was mounted on a backplate machined from a 1 mm Ta sheet that includes the heating wires (Fig. 2a). This minimises the number of connections ensuring optimal thermal contact to the cryostat. The sample was fixed using several Ta strips (see Fig. 2a), and a thin gold foil was placed between the sample and sample plate to improve the thermal contact. The sample was prepared by consecutive cycles of 1 keV $Ne^+$ sputtering at 300 K followed by annealing to 920 K. In every other cycle the sample was reoxidised by exposure to $O_2$ during annealing, which results in the growth of new $Fe_3O_4$(001) surface.[30] For this procedure a MB with a core $O_2$ pressure of $5\times10^{-6}$ mbar was used. To ensure even exposure, the sample was set at an angle of 60° to the beam and moved up and down continuously. After cleaning, a sharp (√2×√2)R45° pattern was observed in LEED, and no signal was observed in the C *1s* region in XPS. The O *1s* and Fe *2p* regions were typical[5] for the clean $Fe_3O_4$(001) surface exhibiting the subsurface cation vacancy (SCV) reconstruction.[31]

$CO_2$ was dosed to a 3.5 mm spot in the centre of the $Fe_3O_4$(001) sample held at 65 K using the MB source described above. A beam reservoir pressure of 0.53 mbar was used, which corresponds to a nominal beam core pressure $p_{MBc} = 2.7\times10^{-8}$ mbar, and an exposure of 1 L in 50 s. The exact orifice–sample distance is determined by measuring the beam spot size on the sample.

TPD experiments were performed with a linear ramp (0.5 K/s). Initial tests revealed a small increase in the residual gas pressure linked to desorption of gases adsorbed on the cryostat when the sample was heated. This effect was removed by stabilizing the temperature of the cryostat at 20 K (by counter heating with an internal heater). Thereafter only $CO_2$ was observed to desorb from the sample during the TPD ramp, so only m/z=18 amu (to monitor for changes in background) and m/z=44 amu were followed during acquisition of the final TPD spectra.

XPS spectra were acquired using the Al Kα anode and a pass energy of 16 eV. Gold, silver and copper foils are mounted on the cryostat support for XPS calibration. Reproducibility in sample position for XPS is achieved using crossed laser beams. The Au foil used to improve thermal contact between sample and mount was also used as a Fermi level reference. UPS spectra were taken with the He II line and a pass energy of 16 eV.

The STM experiments were performed in a separate UHV system with a base pressure $6\times10^{-12}$ mbar using an Omicron LT-STM in constant current mode with electrochemically etched W tips. Here, a synthetic magnetite single crystal was prepared by 1 keV $Ar^+$ sputtering followed by heating to 920 K. Again, every other annealing cycle was performed in a background pressure of $1\times10^{-6}$ mbar $O_2$.

## 4. $CO_2$ ON $Fe_3O_4$(001) – RESULTS

### 4.1 Temperature Programmed Desorption

A series of TPD spectra acquired for different nominal doses are shown in Fig. 3a. For each TPD curve the nominal dose is given in molecules per $cm^2$, deduced from the exposure time and the core intensity of the MB as calculated from the gas reservoir pressure and the geometry. The relationship between the nominal dose and the actual $CO_2$ coverage is derived from Figure 4, and is described later. Selected curves are displayed in an inverted Arrhenius plot [32] in the middle panel (Fig. 3b). The three main desorption peaks below 115 K are labelled as first monolayer (1st ML), second monolayer (2nd ML), and multilayer. Additionally, a non-zero desorption rate is visible between the 1st and 2nd ML peaks (this is clearest in the grey curves in Fig. 3b), and three small desorption peaks related to surface defects are visible at higher temperatures (Fig. 3c). In the following, we discuss these desorption features in turn.

In Fig. 3a the 1st ML peak appears to exhibit zero order desorption kinetics, i.e., desorption rate = ν exp(-$E_d$/RT), where $E_d$ is the desorption energy, T is temperature, R is the gas constant and ν is a pre-exponential constant. However, on close inspection of the inverted Arrhenius plot it is clear that only the TPD curves with coverages from $1.80\times10^{14}$ $cm^{-2}$ to $4.22\times10^{14}$ $cm^{-2}$ have their leading edges aligned, marked with an orange line. From the slope and intercept of this line, one obtains a desorption energy $E_d$=0.4±0.02 eV and prefactor ν=$10^{30\pm1}$ $cm^{-2}s^{-1}$.[33, 34] It is important to note however, that the curves below $1.24\times10^{14}$ $CO_2/cm^2$ have leading edges that shift to higher temperatures with increasing coverage. Together with the slight kink in the tail around 1/T=0.0087 (marked by arrow in Fig. 3b), this suggests that a more complex desorption mechanism occurs in the fractional monolayer regime.

In Fig. 4 we plot the integrated area of each TPD curve versus the nominal dose (determined from the calculated molecular beam pressure and exposure time). The integrated areas are normalised such that a value of 1.0 corresponds to the saturated 1$^{st}$ ML peak, i.e. the area shaded yellow in Figure 3b. The data exhibit a linear dependence, consistent with a constant sticking coefficient. Since the sticking is unity at 65 K (see Fig. S4), we can then apply a linear fit and transform the integrated peak areas into an absolute coverage (shown on the other two axes in terms of $CO_2$ molecules per cm$^2$ and molecules per ($\sqrt{2}\times\sqrt{2}$)R45° unit cell). The peak area measurement is more precise than the nominal dose, partly due to variations in the source pressure caused by the temperature of the room, and partly because the signal/noise from the mass spec is very good. Based on this procedure we find the 1$^{st}$ ML peak saturates at a coverage of $5.25\times10^{14} \pm 0.25\times10^{14}$ $CO_2$/cm$^2$, which corresponds to an areal density of $3.7 \pm 0.18$ $CO_2$ molecules per ($\sqrt{2}\times\sqrt{2}$)R45° unit cell. The estimated error for each dose results from both the error in dose time and variation in the source pressure measurement over time. The resulting error in the absolute coverage is obtained from the linear regression of the linear fit in Fig. 4. The calculated density of 3.7 molecules per unit cell is reasonable because there are 4 Fe cations per ($\sqrt{2}\times\sqrt{2}$)R45° unit cell with which $CO_2$ can interact. Moreover, in what follows we show that a monolayer coverage of 4 $CO_2$ molecules per unit cell is consistent with both XPS measurements (section 4.2) and STM images (section 4.3).

Following the saturation of the 1st ML peak, a small, but non-zero desorption rate is observed to shift to progressively lower temperatures (grey curves in Fig. 3b) until the onset of the 2$^{nd}$ ML peak occurs at 78 K. The coverage at which the 2$^{nd}$ ML begins is $5.88\times10^{14} \pm 0.27\times10^{14}$ $CO_2$/cm$^2$, or $4.14 \pm 0.19$ $CO_2$ molecules per ($\sqrt{2}\times\sqrt{2}$)R45° unit cell. Thus, with increasing $CO_2$ chemical potential it becomes favorable to press additional $CO_2$ molecules into the 1$^{st}$ ML before initiating growth of the 2$^{nd}$ ML. Similar behaviour has been reported previously for the Ar on Pt(111) system.[35]

The 2$^{nd}$ ML peak (red curves) then grows with increasing coverage and saturates at a coverage of $11.4\times10^{14} \pm 0.37\times10^{14}$ $CO_2$/cm$^2$ ($8 \pm 0.26$ $CO_2$/unit cell). The peak maximum is at T = 84 K. Somewhat surprisingly, when more $CO_2$ is dosed, the 2$^{nd}$ ML peak decreases in intensity and a new peak grows in at 88 K (green curves in Fig. 3a). This new peak does not saturate and is therefore clearly due to $CO_2$ multilayers. As expected, the multilayer peak exhibits a zero-order line shape.

Coverages in the transition region between the 2$^{nd}$ ML and multilayer regime are plotted green in Fig. 3. In this transition region, a small variation in coverage causes a significant change in ratio of 2$^{nd}$ ML peak and multilayer, pointing to a kinetic re-organization. In order to test this hypothesis, we compared TPD spectra for different heating rates. Fig. 5 shows TPD traces from a $CO_2$ coverage of $13.0\times10^{14} \pm 0.43\times10^{14}$ $CO_2/cm^2$ desorbing with heating rates of 2.5 K/s and 0.5 K/s, respectively. When the experiment is conducted with a slower ramp rate, where more time is allowed for reconfiguration, more $CO_2$ is transferred into the multilayer (higher-temperature) state. Note that in both cases the total $CO_2$ desorption and the amount specifically within the 1st monolayer peak is the same.

Fig. 3c shows a zoomed region of the temperature range above the 1st ML peak for the TPD curve with coverages $5.42\times10^{14} \pm 0.26$ $CO_2/cm^2$. We assign the three small peaks at temperatures 125 K, 165 K and 195 K (the combined peak area is ≈2% of the first monolayer peak) to adsorption at defects. This assignment is partly based on the STM measurements presented in section 4.3, which reveal that $CO_2$ binds preferentially at surface defects including antiphase domain boundaries (APDBs) and incorporated Fe defects.

**4.2 Photoelectron Spectroscopy**

Figure 6a shows UPS data acquired at normal exit from the clean $Fe_3O_4(001)$ surface and following deposition of $1.2\times10^{15}$ $CO_2/cm^2$. The surface was then heated to 98 K to desorb all but the 1st ML peak (compare Fig. 3a). Figures 6b and 6c show XPS data (O 1$s$ and C 1$s$, respectively) acquired from the as-prepared $Fe_3O_4(001)$ surface and following deposition of $5.5\times10^{14}$ $CO_2/cm^2$ $CO_2$ at 65 K using the MB. Again, this means that only $CO_2$ in the 1st ML peak is present on the surface. The XPS data were acquired at grazing emission (80° off normal).

The clean-surface UPS data appears as reported previously, with a small peak at 0.5 eV and density of states at the Fermi level linked to the octahedrally coordinated Fe cations.[36] Following $CO_2$ adsorption three new peaks are observed at 12.4 eV 10.7 eV, and 6.8 eV assigned as the $1\pi_g$, $3\sigma_g$, and $4\sigma_g$ peaks of $CO_2$, respectively. The 5.6 eV separation of the $1\pi_g$ and $4\sigma_g$ is identical to gas-phase $CO_2$,[37] which supports that $CO_2$ is physisorbed on $Fe_3O_4(001)$.

In XPS, the clean $Fe_3O_4(001)$ surface exhibits a peak at 530.1 eV in O 1$s$, which is asymmetric due to the metallic nature of the oxide.[36] No C 1$s$ signal is observed on the as-prepared surface. Following $CO_2$ adsorption, new peaks appear in the O 1$s$ and C 1$s$ regions at 534.9 eV and 291.3eV, respectively. These positions are similar to those reported on Ni(110) surface, where the $CO_2$ is physisorbed in a linear configuration.[38] C1s spectra of $0.18\times10^{14}$ $CO_2/cm^2$ (see Fig. S5), which corresponds to saturation of the defect peaks plus 1% of the monolayer coverage, reveal a single peak at 291.3 eV, identical to that of the physisorbed monolayer. Since carbonate species typically appear in the range 287-289 eV,[39, 40] we conclude that $CO_2$ adsorbed at defects is also physisorbed.

To further check our assertion that the 1$^{st}$ ML peak in TPD corresponds to four CO$_2$ molecules per ($\sqrt{2}\times\sqrt{2}$)R45° unit cell we compared the C 1s peak area to a monolayer of formate (HCOO) species (not shown), formed by dissociative adsorption of HCOOH on the Fe$_3$O$_4$(001) surface at room temperature.[41] As expected, the C1s peak area from CO$_2$ is twice that of the formate, which has a saturation coverage of two molecules per ($\sqrt{2}\times\sqrt{2}$)R45° unit cell due to its bidentate binding configuration.[41]

### 4.3 Scanning Tunneling Microscopy

To characterize the arrangement of the CO$_2$ molecules on the Fe$_3$O$_4$(001) surface we performed STM experiments. Empty-states images of the as-prepared surface exhibit the characteristic undulating rows of protrusions related to fivefold-coordinated Fe$^{3+}$ cations within a distorted surface layer (Fig. 7a). A structural model of this surface is overlaid in the STM image, and shown in more detail in Fig. 7b. The lattice distortion is caused by an ordered array of cation vacancies and interstitials in the subsurface,[31] and results in a ($\sqrt{2}\times\sqrt{2}$)R45° periodicity. The ($\sqrt{2}\times\sqrt{2}$)R45° unit cell is shaded in the overlay, and indicated by a black square in Fig. 7b. The subsurface tetrahedral Fe (white balls) and surface oxygen atoms (red balls) are not imaged in STM, the latter because there are no O-derived states in the vicinity of $E_F$. The grey shaded area in Fig. 7b highlights the region of the unit cell without a light-grey 2$^{nd}$ layer Fe atom; this is the preferred adsorption site for many metal adatoms and hydrogen.[42-46]

Figure 7c shows an STM image acquired following saturation exposure of 2 Langmuir (L; 1 L = 10$^{-6}$ torr s) CO$_2$ at a nominal sample temperature of 82 K. Note that this temperature is within the 2$^{nd}$ ML desorption peak in Fig. 3a, so only the 1$^{st}$ ML molecules should be present on the surface when the STM experiment is conducted. A further exposure of 1 L produced no discernible effect. Four protrusions per ($\sqrt{2}\times\sqrt{2}$)R45° unit cell are clearly observed, in excellent agreement with the density of CO$_2$ molecules determined from the molecular beam intensity calculations (section 3.1). Interestingly, the protrusions are arranged as alternating bright and dark pairs along the direction of the surface Fe-rows, producing a pattern with the ($\sqrt{2}\times\sqrt{2}$)R45° symmetry of the underlying substrate. The protrusions within each pair are shifted laterally perpendicular to the Fe row direction, as depicted in the schematic shown in Fig. 7d.

The position of the bright and dark pairs relative to the underlying substrate in Fig. 7b and 7d was determined by watching the formation of the overlayer by dosing CO$_2$ directly into STM whilst scanning (see Fig. S6). By aligning the before and after images to surface defects it is possible to assign the location of CO$_2$-related protrusions to the surface Fe row, and determine the position of the bright and dark pairs with respect to the surface reconstruction. Images of the Fe$_3$O$_4$(001) surface with a submonolayer CO$_2$ coverage (Fig. 8) exhibit islands identical to those observed at saturation coverage, with two symmetrically equivalent (mirror) configurations. In the regions in-between, the surface resembles the clean surface, but the Fe rows exhibit a scratchy appearance. Such images are frequently observed when adsorbate molecules are mobile on the time scale of the STM measurement and/or interact with the STM tip.

The very initial stages of adsorption were studied in a further STM experiment where $CO_2$ was dosed at a pressure of $5\times10^{-11}$ mbar on the clean $Fe_3O_4(001)$ surface at 77 K whilst scanning with the STM. Figure 9 shows two representative images selected from a much longer image sequence on the same sample area. The clean surface exhibits several defects, which we identify based on previous work as surface hydroxyl ($O_{surface}H$) groups,[46] an antiphase domain boundary (APDB),[47] and an incorporated Fe defect.[48] The latter defect occurs when the presence of an additional Fe atom in the surface leads to a local lifting of the $(\sqrt{2}\times\sqrt{2})R45°$ reconstruction, and easily are distinguished from $O_{surface}H$ at room temperature by their lack of mobility. The incorporated Fe and APDB defects were recently shown to contain $Fe^{2+}$ cations, and to be active sites for methanol adsorption.[48] Here we observe a preferential adsorption of $CO_2$, with bright protrusions appearing at the position of the defects while scanning with the STM, but no similar events on the defect-free surface in-between. Thus we conclude that $CO_2$, a Lewis acid like methanol, interacts more strongly with the $Fe^{2+}$ sites associated with these defects than with the regular surface.

## 5. DISCUSSION

On the basis of the STM images, quantitative TPD measurements and spectroscopic data presented here it is clear that $CO_2$ adsorbs molecularly on the $Fe_3O_4(001)$ surface. Adsorption occurs initially and most strongly at defects, and subsequently at regular fivefold coordinated $Fe^{3+}$ sites. An ordered structure is formed when the coverage reaches four $CO_2$ molecules per $(\sqrt{2}\times\sqrt{2})R45°$ unit cell, and each molecule is associated with one surface Fe cation. The clear separation of desorption from $Fe^{2+}$ and $Fe^{3+}$-related sites means $CO_2$ can be a useful probe of the relative density of such sites on magnetite surfaces.

For coverages between 1 and 4 molecules per $(\sqrt{2}\times\sqrt{2})R45°$ unit cell the TPD spectra data exhibit aligned leading edges, consistent with zero-order desorption kinetics. Such behaviour has been observed previously in the first monolayer for various molecules [49-53] and results from the coexistence of individual adsorbates (2D gas) and a two-dimensional condensed phase in equilibrium.[50] The chemical potential, and hence the vapour pressure and desorption rate, is defined by the two-phase coexistence, and as long as surface diffusion remains faster than desorption, the desorption rate is independent of coverage.[54] Moreover, the molecule-molecule interaction within the condensed phase must be strong compared to the molecule-substrate interaction, as this ensures that evaporation from the condensed phase into the 2D gas is the rate determining step. In our case, the "condensed" phase is that observed at saturation coverage (Fig. 7c), with four $CO_2$ molecules per $(\sqrt{2}\times\sqrt{2})R45°$ unit cell bound to the surface Fe cations, but exhibiting a zig-zag linked to the inter-molecular interactions. STM images of the submonolayer regime (Fig. 8) exhibit both this structure and scratchy areas linked to fast-diffusing molecules of the 2D gas. The desorption energy of $E_d = 0.4\pm0.02$ eV obtained from the Arrhenius plot in Fig. 3b is similar to that predicted theoretically for $CO_2$ adsorbed at a $Fe^{3+}$ cation on $Fe_3O_4(111)$,[20] and is close to that measured for $CO_2$ physisorbed via the oxygen atoms at $Ti^{4+}$ cations on $TiO_2(110)$ (0.46 eV).[8] However, $CO_2$ does not exhibit zero-order kinetics on $TiO_2(110)$, despite the fact that the adsorption energies are close to that observed here (0.4 eV vs. 0.45 eV respectively). Probably the key difference stems from the surface corrugation. Although the cation-cation distance along the row (0.3 Å) is similar, on $TiO_2(110)$ the rows are separated by bridging oxygen atoms that protrude from the surface. These atoms may prevent the formation of an equilibrium between the 2D gas and the condensed 2D phase, and/or limit diffusion during desorption. Such structure sensitivity is well known in the Xe/Pt system; zero-order kinetics prevail from the flat Pt(111) surface, whereas first-order kinetics dominate on the stepped Pt(997) surface.[53, 55-58] In contrast, $Fe_3O_4(001)$ is a flat surface, as is the only other metal-oxide surface where zero-order kinetics has been observed to date: the ultrathin FeO(111) film grown on Pt(111).[21, 22, 59, 60]

For coverages up to 1 molecule per $(\sqrt{2}\times\sqrt{2})R45°$ unit cell the leading edges of the TPD curves shift to higher temperature with increasing coverage. This is symptomatic of first-order kinetics, and suggests that the "condensed" $CO_2$ phase with 4 molecules per unit cell does not form up to this (surprisingly high) coverage. This raises the possibility that there might in fact be a second ordered phase with the lower density, in which case zero-order kinetics could also result from the coexistence of two different "condensed" structures, as reported by Nagai and Hirashima [52] for the H/Ni(110) system. We were unable to observe a lower-density ordered phase by STM, but this could be due to tip-adsorbate interactions. The kink in the tail of the TPD spectra (arrow in Fig. 3b) suggests that the condensed phase with 4 molecules per unit cell disappears before the last molecules have left the surface.

The STM images of the condensed phase presented in Fig. 7c clearly exhibit a zig-zag orientation along the Fe-row direction. We propose that this results from a quadrupolar interaction i.e. attraction between the C and O atoms of neighboring molecules. Similar behavior was reported on the $TiO_2(110)$ surface, and linked to pairs of $CO_2$ molecules bound to the $Ti^{4+}$ sites through the O atoms that tilt away from each other. The origin of the bright/dark contrast between alternating pairs is harder to explain, but is clearly linked to the underlying surface reconstruction. It is possible that the molecules tilt differently depending on their position relative to the distorted surface layer. On the other hand, we find that that the apparent height depends on bias, suggestive of an electronic origin instead. However, it is impossible to discount that the presence of the STM tip induces tilting in the molecules, so the exact origin must remain speculative at this point.

We now turn our attention to $CO_2$ coverages higher that 4 molecules per $(\sqrt{2}\times\sqrt{2})R45°$ unit cell. After the saturation of the 1st ML peak, a non-zero desorption rate is observed that shifts rapidly to lower desorption temperature with increasing coverage until the second-layer peak appears at 4.5 molecules per $(\sqrt{2}\times\sqrt{2})R45°$ unit cell. This behavior is typical for a compression of the first monolayer.[35, 54] Thus, additional $CO_2$ molecules are squeezed into the ordered structure. The non-integer number of molecules implies there is no one site in the unit cell where an additional $CO_2$ molecule gets accommodated, rather the incorporation must result in displacement of the already adsorbed $CO_2$ away from their favored position above the Fe cation. It seems most likely that the additional $CO_2$ molecule is accommodated within the existing zig-zag chain, which makes sense because the C-O distance between neighboring molecules at a density of 4.5 molecules per unit cell would be shortened to 2.7 Å, akin to the minimum distance in $CO_2$ ice.

Once the compact first layer is formed, a second $CO_2$ layer grows to completion with a similar density as the first layer. However, it is less strongly bound than crystalline $CO_2$ ice (desorption at 84 K vs. 88 K), and therefore most likely has a different structure (most likely planar) influenced by the planar wetting layer. Once the coverage exceeds two layers, however, the 2nd ML peak diminishes and eventually disappears. It is clear from Fig. 5 that the process is kinetically limited, because a slower TPD ramp provides more time for the reorganization to occur. Thus, the two-layer structure formed upon low-temperature adsorption is only metastable. One possible explanation is that the two-layer structure converts to a new, more stable structure during the TPD ramp. In this scenario, the addition of $CO_2$ islands in a third layer would locally induce the phase change, which would then slowly spread throughout the bilayer film. The alternative explanation is that the two-layer structure is unstable against the growth of multilayer islands. The growth of 3D clusters on a wetting layer has been observed for water on Ru(0001), and was linked to the formation of non-ice-like wetting layer structures due to a mixture of hydrogen bonding and a strong interaction with the substrate,[61] similar to Stranski-Krastanov growth. The $CO_2$/$Fe_3O_4$(001) system differs in that the second layer forms completely, and island growth would have to remove $CO_2$ from this structure in a process similar to detwetting. Dewetting in water clusters has been attributed to the additional stabilization of water clusters achieved through hydrogen bonding.[62-64] Since quadrupole-quadrupole interactions dominate the structure of $CO_2$ ice, it seems likely that these interactions would underlie the restructuring observed here.

Finally, XPS measurements reveal that all $CO_2$ is physisorbed on $Fe_3O_4$(001), even at the $Fe^{2+}$ related defects. Chemisorption, including the formation of carbonate species, has been predicted theoretically for an Fe-rich termination of $Fe_3O_4$(111), so it may also occur on Fe-rich terminations of $Fe_3O_4(001)$, which form under very reducing conditions.[65] It is also important to note that OH groups and/or molecular water are always present in the ambient and can play a major role in adsorption. For example, experiments conducted in liquid water reveal bicarbonate $HCO_3$ formation via atmospheric $CO_2$ on $TiO_2(110)$.[66] No such reaction happens under UHV conditions,[49] indicating a significant pressure gap.

## 6. SUMMARY

This paper describes an investigation of $CO_2$ adsorption on $Fe_3O_4(001)$ conducted using a newly-constructed vacuum system optimized for the study of metal-oxide single crystals. The combination of a special sample mount and molecular-beam dosing allows quantitative TPD experiments to be performed on a natural $Fe_3O_4(001)$ single crystal with high precision and reproducibility. The TPD data, together with photoelectron spectroscopy and scanning tunneling microscopy images, allow developing of a full picture of the behaviour of $CO_2$ on this surface. Adsorption occurs initially at defects, and subsequently at regular $Fe^{3+}$ sites in a 2D gas phase. Above a coverage of $\approx 1$ molecule per unit cell an ordered phase forms in which $CO_2$ is bound to surface iron cations and other $CO_2$ molecules via quadrupole-quadrupole interactions. Coexistence between these two phases leads to zero-order desorption kinetics. When all cations are occupied, additional molecules can still be incorporated into the first layer until it reaches a nearest neighbour distance similar to $CO_2$ ice. A complete second monolayer grows to completion on top of the first, but is ultimately unstable against the formation an ice-like structure, or multilayer islands.

## SUPPLEMENTARY INFORMATION

See supplementary information for further details about the molecular beam design, including calculations of the molecular beam intensity, an SEM image of the orifice, and measurements of the molecular beam profile using the MBM. Also, we show the sticking of $CO_2$ at 60 K versus 300 K, C1s XPS at low coverage, and further STM images of the same area before and after saturation $CO_2$ exposure.

## ACKNOWLEDGEMENTS

G.S.P., R.B., O.G., J.H., and J.P. acknowledge funding from the Austrian Science Fund START prize Y 847-N20 and project number P24925-N20. O.G. acknowledges a stipend from the Vienna University of Technology and the Austrian Science Fund as part of the doctoral college SOLIDS4FUN (W1243). U.D. and J.P. acknowledge support by the European Research Council (Advanced Grant "OxideSurfaces"). M.S. was supported by the Austrian Science Fund (FWF) within SFB F45 "FOXSI". We would like to thank Rainer Gärtner and Herbert Schmidt from the workshop at the TU Wien, as well as Manfred Bickel and Martin Leichtfried for much work during the construction and commissioning of the new vacuum system.

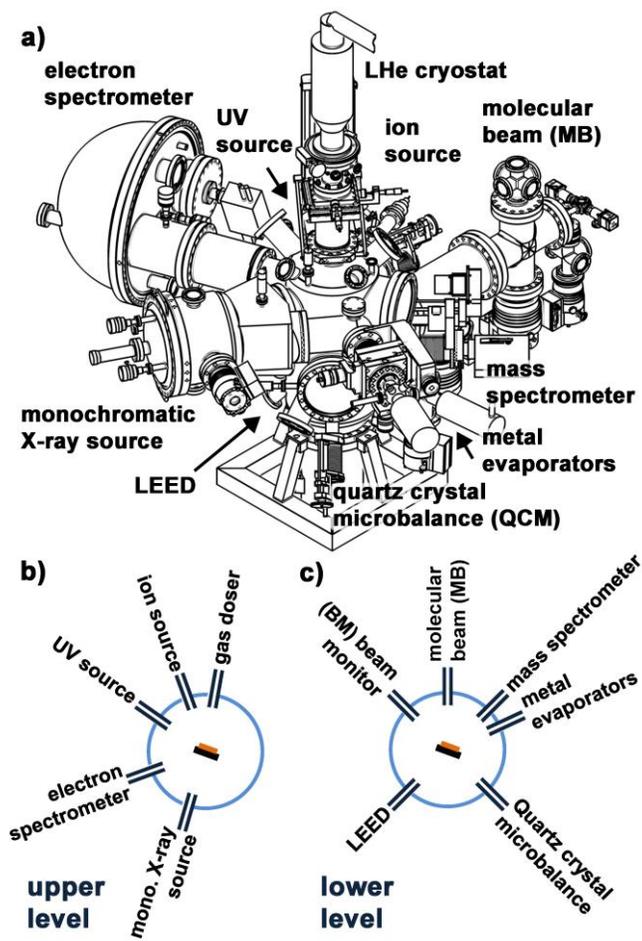

**Fig. 1: a)** Isometrical view of the newly-designed vacuum system utilized in this work. b) Scheme of upper level where an electron spectrometer provides the basis for XPS, UPS, and LEIS experiments. c) Scheme of lower level primarily used for TPD.

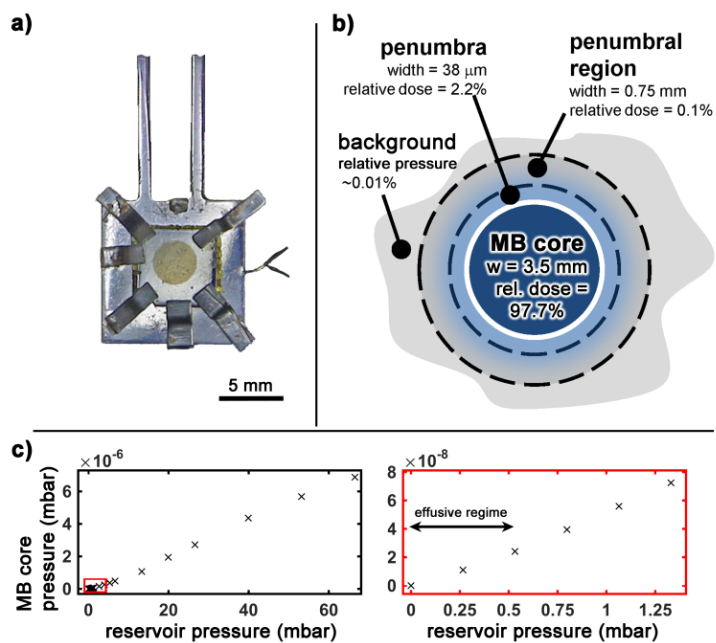

**Fig. 2: The Molecular Beam – design and characterization.** a) The $Fe_3O_4$(001) single crystal mounted on a polished backplate/heating wire machined from a 1 mm Ta sheet. The yellowish circular spot on the sample was created by condensing a thick film of water ice using the MB source. b) Scheme illustrating the calculated MB profile in which 97.7% of the molecules reside within the beam core. c) Plot of the measured MB core pressure versus reservoir pressure showing the linear dependence expected in the effusive regime. The right panel contains a zoomed view of the low intensity data from the left panel.

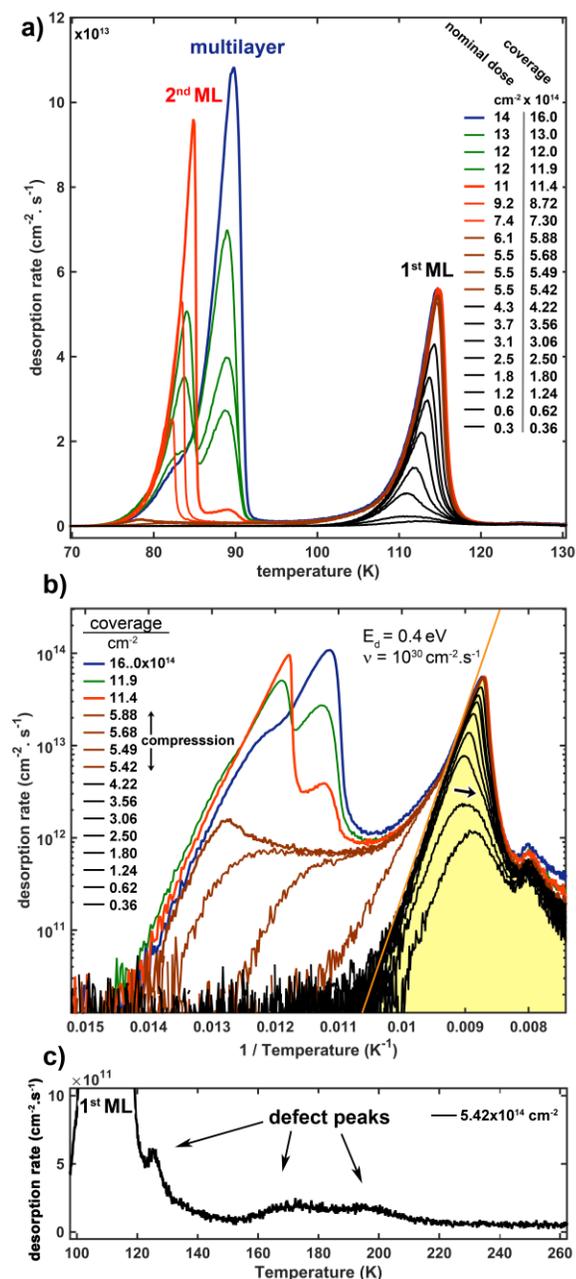

**Fig. 3**: Temperature programmed desorption spectra for $CO_2$ on the $Fe_3O_4(001)$ surface. a) TPD spectra for $CO_2$ adsorbed at 65 K on $Fe_3O_4(001)$ performed with a linear ramp of $\beta=0.5$ K/s. The curves are labelled by the nominal dose (molecules/cm$^2$), and the coverage deduced from the linear fit in Figure 4. b) Inverted Arrhenius plot showing selected TPD curves from panel a. c) Detail of the TPD curve acquired for a coverage of $5.42\times10^{14}$ $CO_2/cm^2$ in the temperature range (110-270 K) showing peaks assigned to desorption from surface defects.

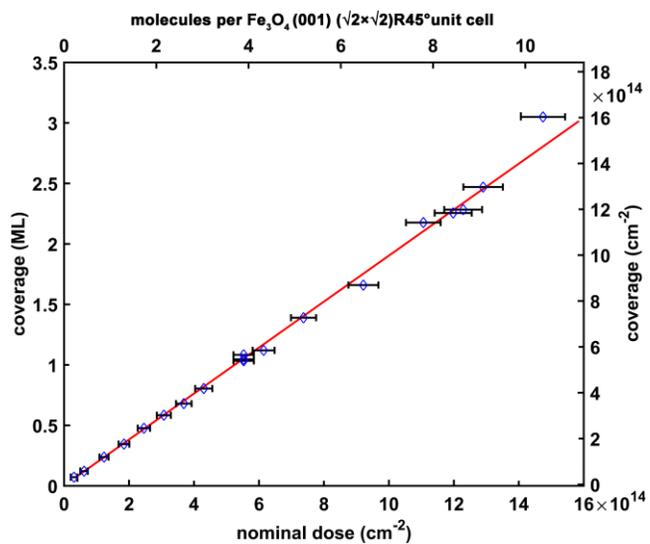

**Fig. 4: Plot of the measured coverage vs. nominal dose, calculated from the molecular beam parameters.** The measured coverage is normalized to the area of the 1st ML peak in TPD. The right-hand y-axis is calculated from a linear fit to the data (see main text).

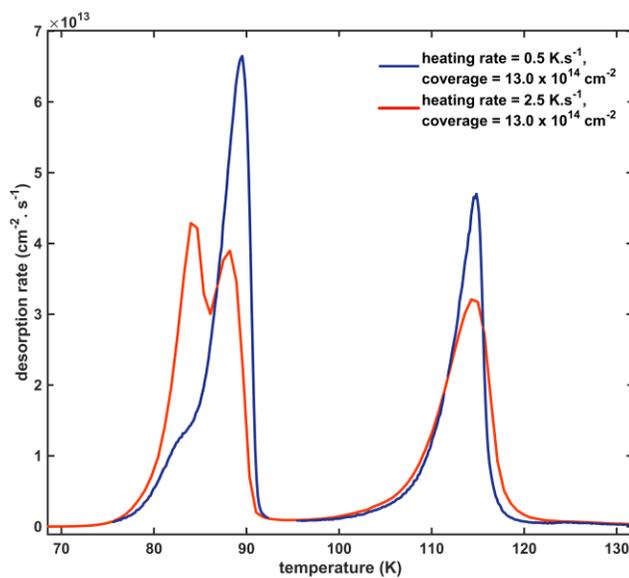

**Fig. 5: TPD for a CO$_2$ coverage of 13.0×10$^{14}$ CO$_2$/cm$^2$ (9.14 molecules per ($\sqrt{2}\times\sqrt{2}$)R45° unit cell) acquired with different heating rates.** The red curve, acquired at 2.5 K/s exhibits both 2$^{nd}$ ML and multilayer desorption peaks. The blue curve, acquired with a ramp of 0.5 K/s for the same initial coverage, features almost exclusively multilayer desorption.

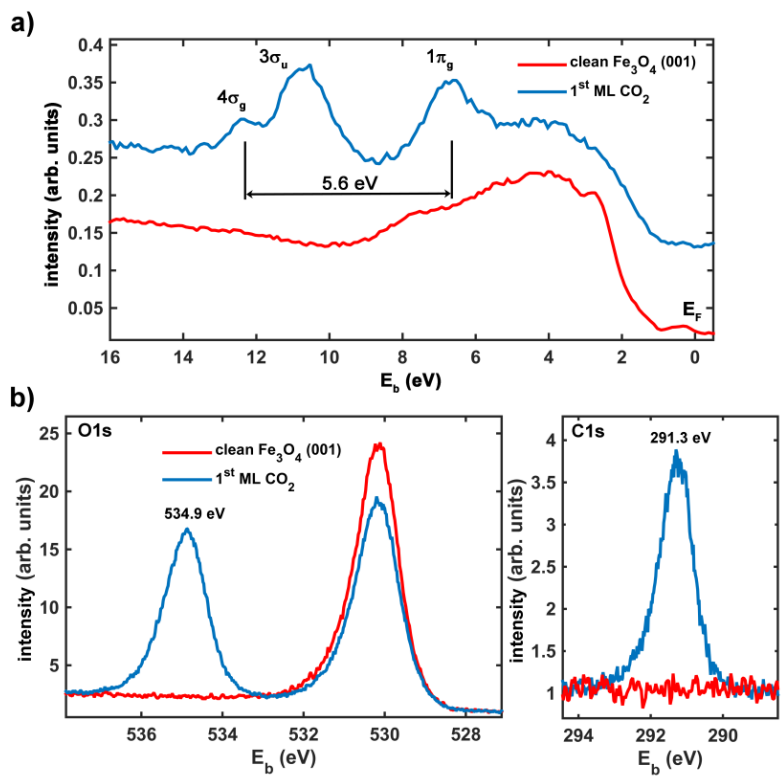

**Fig. 6**: Photoelectron spectroscopy data for $CO_2$ adsorbed on $Fe_3O_4$(001) in the monolayer regime. a) UPS spectra for the clean and $CO_2$ covered $Fe_3O_4$(001) surface acquired with a photon energy of 40.3 eV (He II). b) XPS (O1s and C1s) of adsorbed $CO_2$ on $Fe_3O_4$ (001) measured with Al K$\alpha$ radiation.

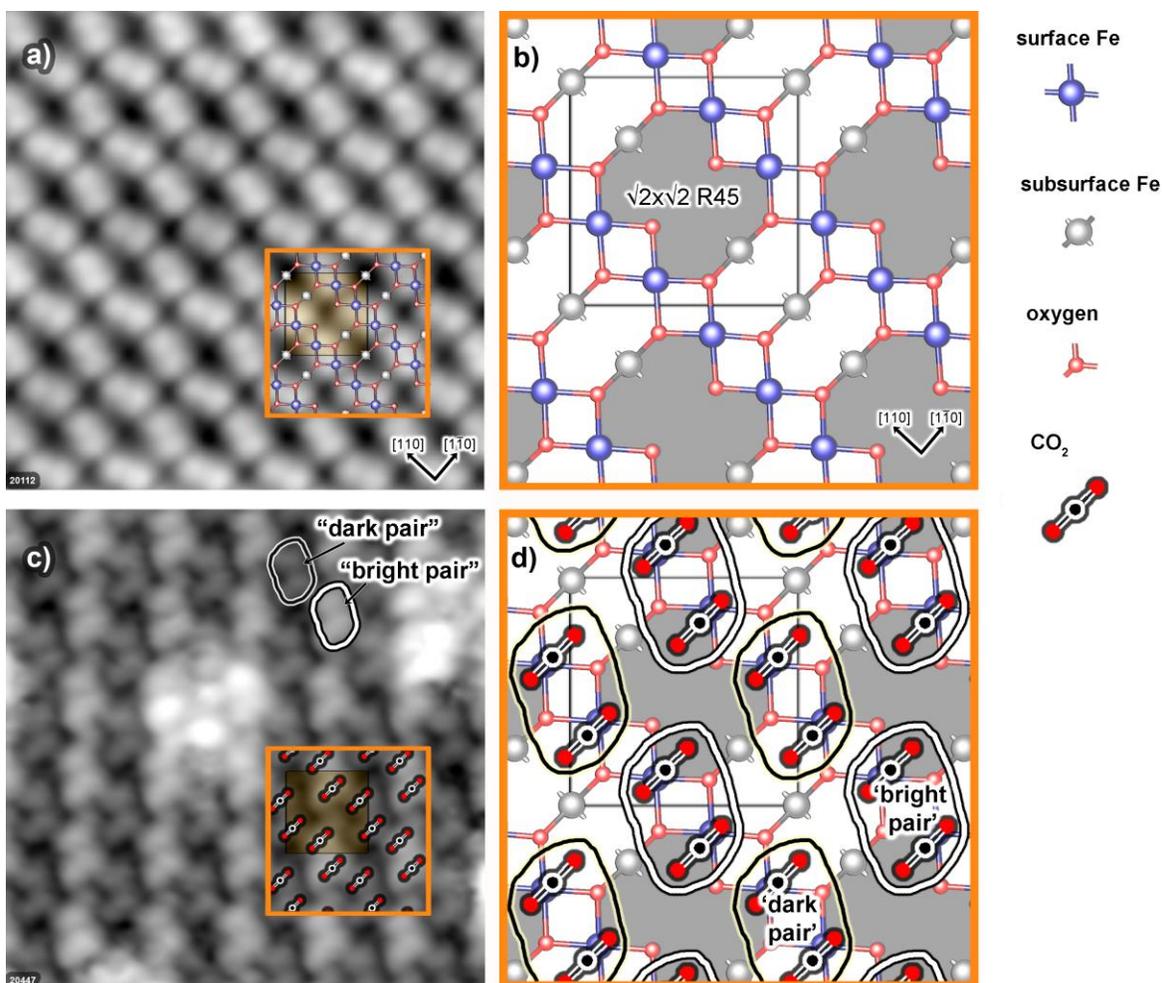

**Fig. 7:** a) STM image (5×5 nm$^2$, $V_{sample}$ = +1.0 V, $I_{tunnel}$ = 30 pA) of the clean Fe$_3$O$_4$(001)-($\sqrt{2}\times\sqrt{2}$)R45° surface measured at 78 K. b) Ball-and-stick model of the clean Fe$_3$O$_4$(001) surface. Surface Fe cations (blue balls) are fivefold coordinated to oxygen (red balls). Grey balls represent subsurface Fe cations (tetrahedral coordination). c) STM image (5×5 nm$^2$, $V_{sample}$ = +0.8 V, $I_{tunnel}$ = 30 pA) of CO$_2$ adsorbed on Fe$_3$O$_4$(001) following saturation exposure at 84 K. A coverage of 4 molecules per unit cell corresponds to the 1$^{st}$ ML TPD peak in Fig. 2. d) Schematic model of CO$_2$ in the 1$^{st}$ ML based on the STM data. The alignment of the bright and dark pairs with respect to the surface reconstruction is based on the STM images shown in Fig. S5.

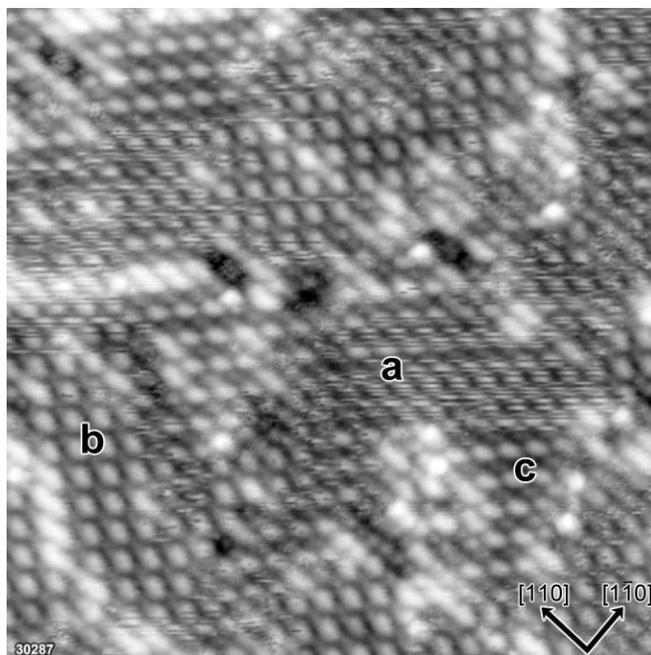

**Fig .8:** STM image (20×20 nm$^2$, $V_{sample}$ = +1.0 V, $I_{tunnel}$ = 30 pA, T = 77 K) showing a submonolayer coverage of $CO_2$ adsorbed on $Fe_3O_4$(001). The image exhibits areas (e.g. patch a) that resemble the clean surface, but appear scratchy due to the presence of fast moving adsorbates within the 2D gas. Patches "b" and "c" have a similar structure to the complete monolayer (shown in Fig. 7), but are mirrored about the [110] direction.

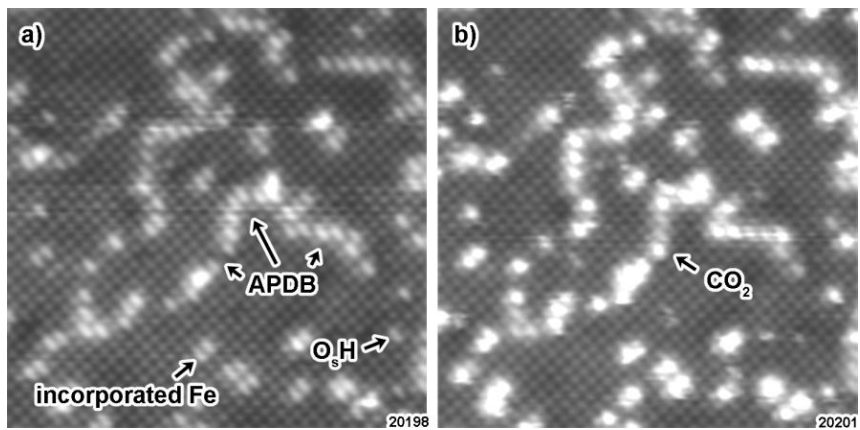

**Fig. 9**: STM images (22×22 nm$^2$, $V_{sample}$ = +1.0 V, $I_{tunnel}$ = 50 pA, T = 77 K) acquired on the same area while exposing the clean Fe$_3$O$_4$(001) surface to CO$_2$ at 5×10$^{-11}$ mbar. a) Prior to introducing the CO$_2$ the surface exhibits an antiphase domain boundary (APDB) defect, several incorporated Fe defects, and surface OH groups. (b) After 8 minutes new protrusions related to CO$_2$ appear at the location of the defects, consistent with stronger adsorption at defects than on the regular surface.

TITLE GOES HERE: UNOFFICIAL WORD TEMPLATE FOR APS

Supplementary Information for:

# A Multi-Technique Study of CO$_2$ Adsorption on Fe$_3$O$_4$ Magnetite

Jiri Pavelec, Jan Hulva, Daniel Halwidl, Roland Bliem, Oscar Gamba, Zdenek Jakub, Florian Brunbauer, Michael Schmid, Ulrike Diebold and Gareth S Parkinson

**Description of molecular beam**

Here we explain how the molecular beam (MB) core pressure at the sample position is calculated. A complete description is contained within the Masters thesis of Daniel Halwidl, which was recently published as part of a book series [1]. A schematic of the MB is shown in Fig. S1. The MB consists of reservoir filled with a gas of pressure $p_r$. The gas expands through an orifice and passes two apertures, denoted separation and exit, on the path to the sample of length L. The separation aperture works as a skimmer and connects the first and second stage of differential pumping. The exit aperture essentially determines the profile of the beam on the sample. For our MB we wished to maintain a Maxwell-Boltzmann gas distribution through the whole gas expansion (i.e. to remain in the effusive regime) and to achieve as close to a top-hat spatial distribution at the sample position as possible. The first condition is met by ensuring molecular flow of gas during the expansion, which means keeping the Knudsen number above 1. The Knudsen number, in this case, is the ratio of the mean free path length to the orifice diameter. In this work, the $CO_2$ reservoir pressure was set to $p_r$ = 0.533 mbar, which corresponds to a Knudsen number of 2.18 at orifice. The second requirement is achieved by using a small orifice, which works almost as a point source, and by placing the exit aperture as close to the sample as is reasonable. The orifice is shown in Fig S1. It was prepared by laser drilling in a 20 µm stainless steel foil, and imaged using a calibrated scanning electron microscope (SEM). The effective diameter of $d_o$ = 37.9 µm was deduced from the area labeled on Fig. S1b). Note that the exact shape of the orifice is unimportant if L >> $d_o$.

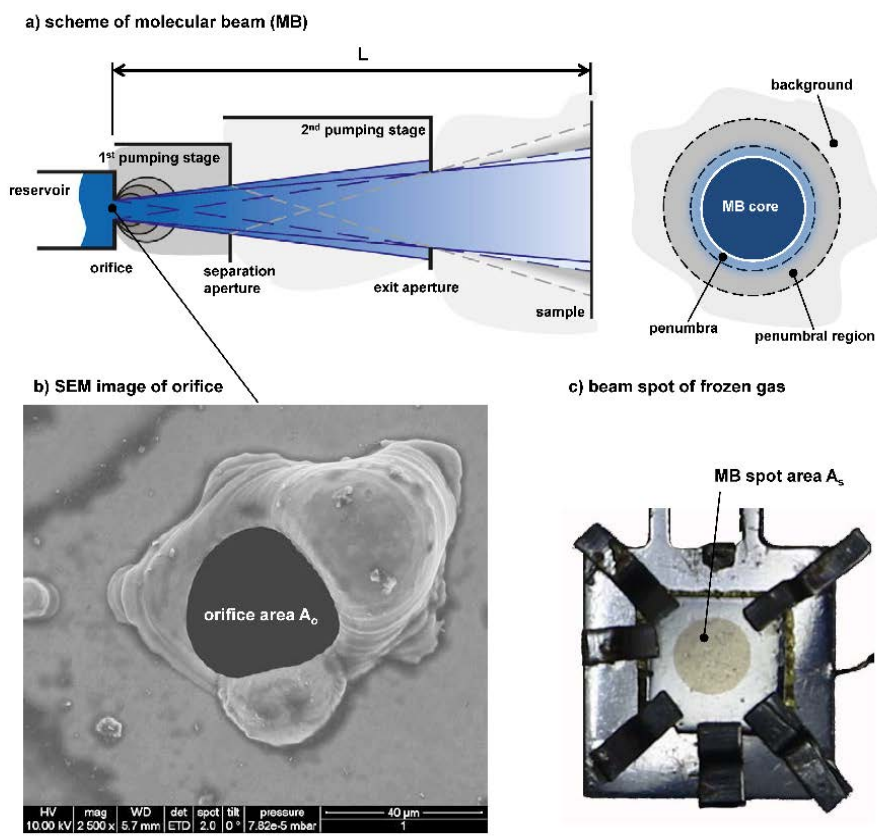

**Fig. S1: a)** Schematic of the molecular beam. **b)** SEM image of the orifice. **c)** Photograph of the mounted sample on which a thick layer of ice was deposited.

The distance L was deduced by condensing a layer of visible ice ($H_2O$) onto the sample (Fig. 1c), and measuring the spot diameter $d_s$. The distance follows from equation 1

$$L = L_a \frac{d_s}{d_a},$$

where $L_a$=51.0 mm is the distance between the orifice and exit aperture, and $d_a$=2.0 mm is the diameter of the exit aperture.

To calculate the MB core intensity the effusive flow of gas from a thin-walled orifice is considered. The intensity in the forward direction for orifice-to-sample distances large compared to the orifice diameter is

$$I(0) = \frac{n_r \bar{v} \sigma}{4\pi} = \frac{p_r}{k_b T} \cdot \frac{\bar{v} \sigma}{4\pi}, \quad I(0) = \text{particles sr}^{-1}.\text{s}^{-1},$$

where $n_r$ is the number density of the gas in the reservoir, $p_r$ is the reservoir pressure, $\bar{v}$ is the average particle velocity, T is the absolute temperature of the gas and $\sigma$ is the orifice area (Scoles, 1988). Hence the intensity (particles per unit area and unit time) in the MB core at a sample in distance L is

$$I = \frac{p_r}{k_b T} \cdot \frac{\bar{v}}{4\pi} \cdot d\Omega = \frac{p_r}{k_b T} \cdot \frac{\bar{v} d_o^2}{16} \cdot \frac{1}{L^2}$$

This intensity is equivalent to the pressure

$$\tilde{p} = I \frac{4 k_b T}{\tilde{v}} = \frac{1}{4} p_r \frac{d_o^2}{L^2}$$

following from the relation between wall collision rate $J_N$ (equivalent to intensity) and pressure p of a gas:

$$J_N = \frac{n\bar{v}}{4} = \frac{p\bar{v}}{4 k_b T}, \quad I(0) = \text{particles sr}^{-1}.\text{s}^{-1},$$

It is important to note that the previous equations are only strictly valid in the molecular flow limit, $Kn \gg 1$. The increasing conductance of the orifice towards the transition flow regime ($10000 > Kn > 1$) has to be considered for the present Knudsen number of 2.18. The ratio of the pressure dependent conductance to the molecular orifice conductance, $\gamma(Kn)$, is shown in Figure 2. Therefore the final expression for the MB pressure is

$$\tilde{p}_{MBc} = \frac{1}{4} p_r \cdot \gamma(Kn_O) \cdot \frac{d_o^2}{L^2}$$

For $CO_2$, $p_r$=0.533 mbar, Kn=2.17, $\gamma(2.17)$ = 1.0252, $d_o$=37.9 µm, L=85.6 mm, and T=300K. This gives $p_{MBc}$ = 2.68×10$^{-8}$ mbar.

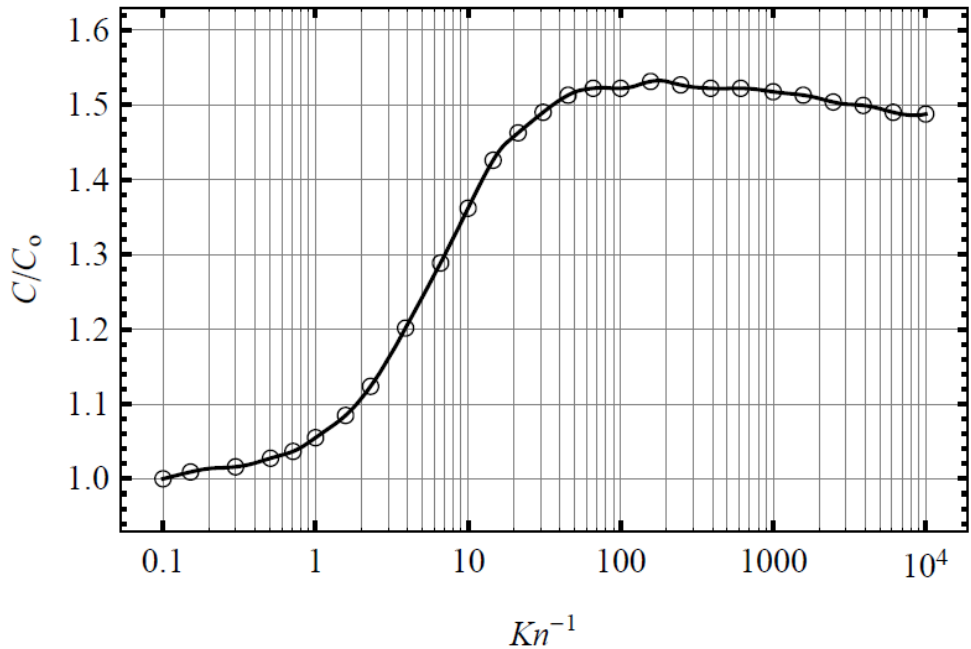

**Fig. S2:** Interpolation (line) of the experimental data (circles) for the ratio $C/C_O$ as a function of the inverse Knudsen number from [(Jitschin, Ronzheimer, & Khodabakhshi, 1999), Fig.2].

**Beam Profile Measurements**

To record the profile shown in Fig. S3, the whole beam monitor assembly is moved around the sample position using an x-y manipulator. The beam profiles shown in Figs. S3a and S3b were measured for Ar, and reveal the beam being close to a top-hat shape, with the intended width of 3.5 mm. The measured edge width (Fig. S3a) of 0.5 mm (i.e. the size of the BM orifice) is consistent with a sharp penumbra. Moving the BM forward and backwards shows that deviations from a top-hat profile visible at the right side of Fig. S3a are due to ionic pumping by the ion gauge.[2]

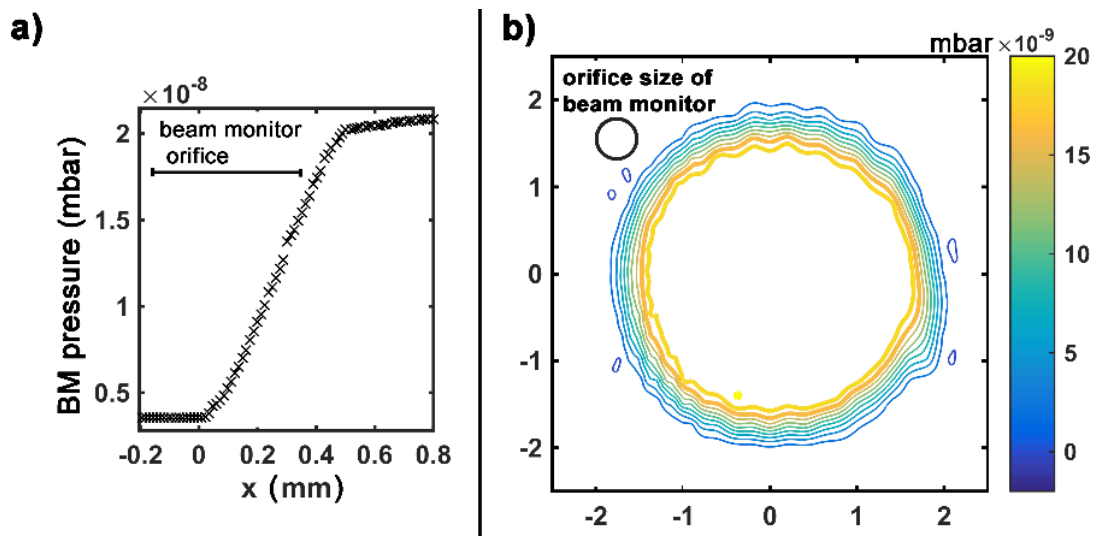

**Fig. S3: a)** Profile of the MB penumbra as measured by the BM. **b)** Measured MB profile plotted as a contour plot.

## CO$_2$ Sticking Coefficient Measurement

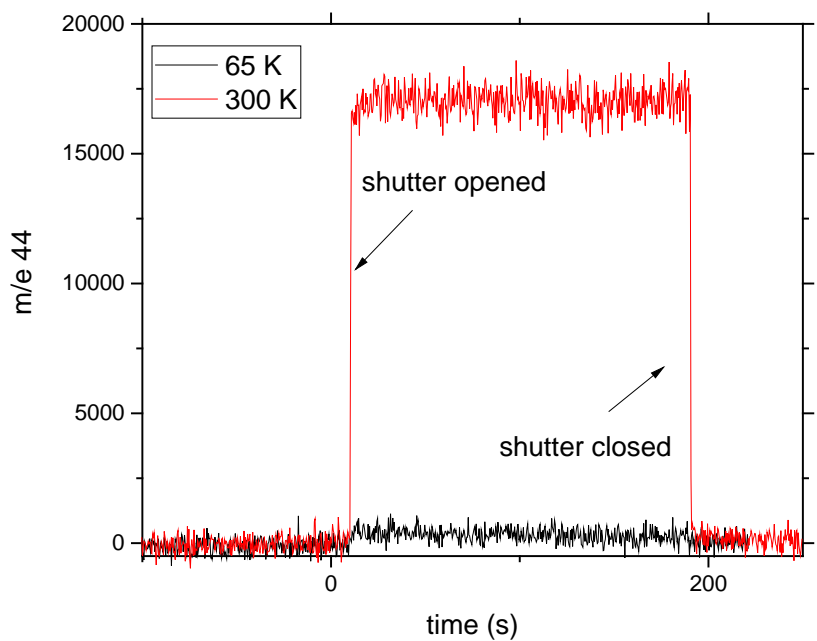

**Figure S4:** Comparison of the mass spectrometer response when the as-prepared Fe$_3$O$_4$(001) sample is exposed to the CO$_2$ molecular beam at 65 K (black) and 300 K (red) in a line of sight geometry. Figure S4 shows that all CO$_2$ sticks to the sample held at 65 K.

## XPS of Defect-Bound $CO_2$

To measure the $CO_2$ bound at $Fe^{2+}$ bound defects we deposited 0.03 ML ($1.8 \times 10^{13}$ $CO_2/cm^2$) onto the as-prepared $Fe_3O_4$(001) surface at 65 K. This saturates the defect peaks (0.02 ML) and places 0.01 ML within the physisorbed first monolayer TPD peak. The C1s spectrum of this surface exhibits a single peak at 291.3 eV, indicating that all $CO_2$ molecules including those bound at defects, are physisorbed.

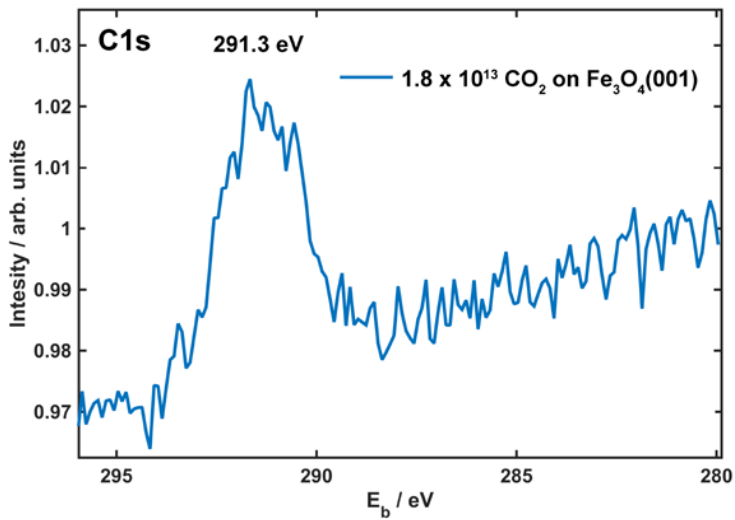

**Fig. S5:** Photoelectron spectroscopy data of C1s region where 0.03 ML of $CO_2$ was adsorbed on $Fe_3O_4$(001) and measured with Al K$\alpha$ radiation. To increase the count rate the pass energy was set to 100 eV.

## STM

The position of the bright and dark pairs relative to the underlying substrate in Fig. 7b and 7d was determined by watching the formation of the overlayer. To achieve this $CO_2$ was dosed directly into STM whilst scanning. In figs. 7e and 7f we show images acquired on the same sample area before and after adsorption of the $CO_2$ monolayer. The resolution of the clean surface is not ideal under these conditions (the individual Fe atoms are not resolved within the row) but the undulation of the rows is visible. Orange ovals mark three surface defects that have already adsorbed $CO_2$. By aligning the before and after images to these markers it is possible to assign the location of $CO_2$-related protrusions to the surface Fe row, and determine the position of the bright and dark pairs with respect to the surface reconstruction (see Fig. 7d).

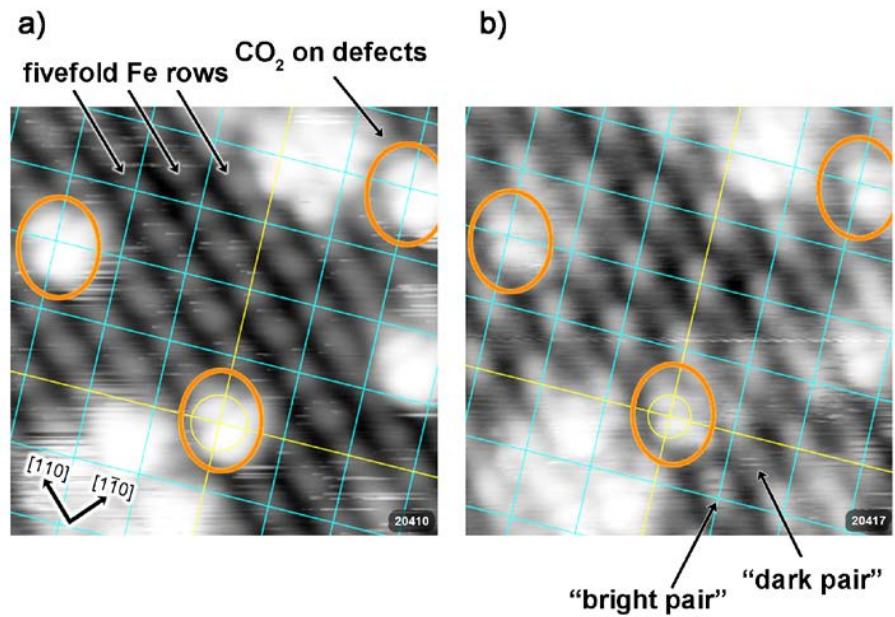

**Fig. S6:** STM images (5.5×5.5nm$^2$, $V_{sample}$ = +1.0 V, $I_{tunnel}$ = 30 pA) of same area before and after dosing saturation coverage at 84 K. $CO_2$ is adsorbed on defects, which are used as markers for the reference grid in both images. These results suggest that the $CO_2$ molecules are adsorbed at the surface Fe atoms at this coverage.